%% file: main.tex
\documentclass[onefignum,onetabnum]{siamart190516}

\usepackage{graphicx}
\usepackage{color}
\usepackage{booktabs}
\usepackage{xspace}
\usepackage{amssymb}
\usepackage{url}

\ifpdf
  \DeclareGraphicsExtensions{.eps,.pdf,.png,.jpg}
\else
  \DeclareGraphicsExtensions{.eps}
\fi

\headers{Compressed Basis GMRES on High Performance GPUs}{J. I. Aliaga et al.}

\title{Compressed Basis GMRES on High Performance GPUs}

\author{%
Jos\'e I. Aliaga\thanks{Universitat Jaume~I, Spain (\email{aliaga@uji.es})}
\and
Hartwig Anzt\thanks{Karlsruhe Institute of Technology, Germany; and
              Innovative Computing Laboratory, University of
              Tennessee at Knoxville, USA
              (\email{hartwig.anzt@kit.edu}).}
\and
Thomas Gr\"{u}tzmacher\thanks{Karlsruhe Institute of Technology, Germany
              (\email{thomas.gruetzmacher@kit.edu})}
\and
Enrique~S.~Quintana-Ort\'{\i}\thanks{
              Universitat Polit\`ecnica de Val\`encia, Spain (\email{quintana@disca.upv.es})}%
\and 
Andr\'es E. Tom\'as\thanks{
              Universitat Jaume~I, Spain; and 
              Universitat de Val\`encia, Spain
              (\email{tomasan@uji.es})} 
}

\usepackage{amsopn}
\newcommand{\spmv}{\textsc{SpMV}\xspace}
\newcommand{\gemv}{\textsc{GeMV}\xspace}
\newcommand{\gko}{\textsc{Ginkgo}\xspace}
\newcommand{\csr}{\textsc{csr}\xspace}

\newcommand{\nxn}{{n \times n}}
\newcommand{\Rnn}{{\mathbb{R}}^{\nxn}}
\newcommand{\Rn}{{\mathbb{R}}^{n}}
\newcommand{\nref}[1]{(\ref{#1})}
\renewcommand{\gko}{Ginkgo}

\def\hyph{-\penalty0\hskip0pt\relax}

\ifpdf
\hypersetup{
  pdftitle={GMRES Solvers with Reduced-Precision Orthogonal Basis for Graphics Processors},
  pdfauthor={%
J. I. Aliaga, 
H. Anzt, 
T. Gr\"{u}tzmacher, 
E. S. Quintana-Ort\'{\i},
A. Tom\'as%
}
}
\fi

\begin{document}

\maketitle

\begin{abstract}
\input{s0-abstract}
\end{abstract}



\input{body}

\section*{Acknowledgments}
Jos\'e I. Aliaga, Enrique~S.~Quintana-Ort\'{\i} and Andr\'es E. Tom\'as
were supported by the EU H2020 project 732631 ``OPRECOMP. Open Transprecision Computing'' and the \emph{MINECO} (Spain) project TIN2017-82972-R. Hartwig Anzt and Thomas Gr\"utzmacher were supported by the ``Impuls und Vernetzungsfond'' of the Helmholtz 
Association under grant VH-NG-1241, and the US Exascale Computing Project 
(17-SC-20-SC), a collaborative effort of the U.S. Department of Energy Office 
of Science and the National Nuclear Security Administration.

\bibliographystyle{siamplain}
\bibliography{biblio}
\end{document}

%% file: s0-abstract.tex
Krylov methods provide a fast and highly parallel numerical tool for the iterative solution of many large-scale sparse linear systems. To a large extent, the performance of practical realizations of these methods is constrained by the communication bandwidth in all current computer architectures, motivating the recent investigation of sophisticated techniques to avoid, reduce, and/or hide the message-passing costs (in distributed platforms) and the memory accesses (in all architectures).

This paper introduces a new communication-reduction strategy for the (Krylov) GMRES solver that advocates for decoupling the storage format (i.e., the data representation in memory) of the orthogonal basis from the arithmetic precision that is employed during the operations with that basis. Given that the execution time of the GMRES solver is largely determined by the memory access, the datatype transforms can be mostly hidden, resulting in the acceleration of the iterative step via a lower volume of bits being retrieved from memory. Together with the special properties of the orthonormal basis (whose elements are all bounded by 1), this paves the road toward the aggressive customization of the storage format, which includes some floating point as well as fixed point formats with little impact on the convergence of the iterative process.

We develop a high performance implementation of the ``compressed basis GMRES'' solver in the Ginkgo sparse linear algebra library and using a large set of test problems from the SuiteSparse matrix collection we demonstrate robustness and performance advantages on a modern NVIDIA V100 GPU of up to 50\%  over the standard GMRES solver that stores all data in IEEE double precision.

%% file: body.tex
\input{s1-intro}

\input{s2b-relatedwork}

\input{s2-alg}

\input{s3-newalg}
\input{s4-impl}

\input{s5-experiments}

\input{s6-conclusions}

%% file: s1-intro.tex
\section{Introduction} 

Krylov solvers enhanced with some type of sophisticated preconditioning technique nowadays compound a popular approach for the iterative solution of large and sparse linear systems~\cite{Saa03}. 
In particular,  preconditioned Krylov solvers are often preferred over their direct counterparts
for the solution of discretized high-dimensional problems (e.g., 3D problems), where a factorization-based direct solver based would incur significant fill-in~\cite{Dav06,Saa03}.
Krylov solvers are also widely appealing for massively-parallel architectures (e.g., graphics processing units, or GPUs) 
due to their superior scalability.

At a high level, Krylov methods span a Krylov subspace by generating a sequence of orthogonal (Krylov) search directions (starting with the normalized residual and computing each new search direction via the multiplication of the sparse coefficient matrix with the previous direction); the orthogonalization of the resulting search direction against previous search directions; and the optimization of the solution approximation in the extended Krylov subspace~\cite{Saa03}.
Each iteration step is usually composed of a sparse matrix-vector product (\spmv); an orthogonalization routine; and several vector operations that compute the new search directions, update the solution approximation, and estimate the norm of the residual~\cite{Saa03}. 

The numerical operations (kernels) appearing in Krylov methods are well-suited for parallelization. 
Unfortunately, most of these kernels, including \spmv, are me\-m\-ory-bound on 
virtually all modern processor architectures~\cite{Hor14}. As a result, many generic as well as hardware-specific optimization efforts for Krylov methods have focused on avoiding, reducing, or hiding (i.e., overlapping with computation) 
the communication/memory accesses of the algorithm.
Some optimization techniques targeting the communication overhead include the following:
\begin{list}{--}{}
\itemsep=0pt\parskip=0pt
\item The design of specialized (i.e., application-specific) sparse matrix data layouts that restrict the indexing information (overhead) and/or
      improve data locality when accessing the contents of the sparse coefficient matrix~\cite{Saa03}.
\item The reorganization of the operations inside the body of the Krylov solver that trades off reduced communication for an increase of computation per iteration, possibly also at the cost
      of introducing numerical instabilities that may result in slower convergence of the iteration;
      see, e.g.,~\cite{COOLS201916} and the references therein.
\item The reformulation of the solver as an iterative refinement scheme 
      combined with the use of mixed precision for the storage of and arithmetic operations with the sparse 
      coefficient matrix~\cite{Higham:2002:ASN}.
\item The utilization of adaptive-precision schemes for memory-bound preconditioners~\cite{Anz19b}. 
\end{list}

In this paper we also address the communication costs of Krylov methods, focusing on
the Generalized Minimal residual (GMRES) algorithm, a Krylov solver for general 
linear systems that 
explicitly maintains the complete set of Krylov search directions instead of 
relying on short recurrences (as many other Krylov solvers do)~\cite{Saa03}. 
Orthogonally to all previous communication optimization efforts, 
our optimized variant of the GMRES algorithm reduces communication in the access to
the orthogonal basis during the iteration loop body. In more detail, 
our GMRES algorithm decouples the memory storage format from the arithmetic precision, and stores the Krylov search directions in a compact ``reduced precision'' format. This radically diminishes the memory access volume during the orthogonalization, while not affecting the convergence rate of the solver, yielding notable performance improvements.
Concretely, we make the following contributions in our paper:
\begin{list}{--}{}
\itemsep=0pt\parskip=0pt
\item We propose to decouple the memory storage format from the arithmetic precision to maintain the Krylov basis in reduced precision in memory while performing all arithmetic operations using full, hardware-supported \textsc{ieee} 64-bit double precision (DP).
\item We analyze the benefits that result from casting the orthogonal basis into different compact storage formats, including the natural \textsc{ieee} 32-bit single precision (SP) and 16-bit half precision (HP) as well as some other non-\textsc{ieee} fixed point-based alternatives enhanced with vector-wise normalization.
\item We provide strong practical evidence of the advantage of our approach by developing a high performance realization of the solver for modern NVIDIA's V100 GPUs and testing it using a considerable number of
large-scale problems from the SuiteSparse matrix collection (\url{https://sparse.tamu.edu/}).
\item We integrate the mixed precision GMRES algorithm into in the Ginkgo\footnote{\url{https://ginkgo-project.github.io}} sparse linear algebra library.
\item We combine our implementation with a high performance realization of an adaptive precision block-Jacobi preconditioner that adjusts the memory format for the distinct diagonal blocks to the numerical properties.
\end{list}

The rest of the paper is organized as follows. In~\Cref{sec:related} we list related work in the direction of mixed precision Krylov solvers. In~\Cref{sec:alg} we briefly recall the GMRES algorithm before motivating in~\Cref{sec:newalg} the \textit{compressed basis GMRES (CB-GMRES)} storing the orthonormal basis in reduced precision. In~\Cref{sec:implementation} we provide details about how we decouple the memory precision from the arithmetic precision, and how we realize the implementation of CB-GMRES in \gko. The experimental evaluation of the CB-GMRES implementation is presented in~\Cref{sec:experiments}, assessing accuracy, convergence, performance, and flexibility of the developed algorithm. We conclude in~\Cref{sec:conclusion} with a summary of the findings and ideas for future research.

%% file: s2b-relatedwork.tex
\section{Related work}
\label{sec:related}
The potential of using lower precision in different components of a Krylov solver has been previously investigated for both Lanczos-based (short-term recurrence) and Arnoldi-based (long-term recurrence) algorithms and the associated methods for solving linear systems of equations. 

From the theoretical point of view, most of those works are based on rounding theory for Krylov solvers running in finite precision. Among the most relevant results are those by Paige~\cite{Paige1980}, who derived distinct relations between the loss of orthogonality and other important quantities in finite precision Lanczos. Greenbaum extended these results to prove 
backward stability for the CG method in finite precision~\cite{Greenbaum89}.
She also derived theoretical bounds for the maximum attainable accuracy in finite precision for CG, BiCG, BiCGSTAB, and other Lanczos-based methods~\cite{Greenbaum97}. 
Carson~\cite{Carson2015} extended these results to 
$s$-step Lanczos/CG variants, deducing that an $s$-step Lanczos in finite precision behaves like a classical Lanczos run in lower ``effective'' precision, where this ``effective'' precision depends on the conditioning of the polynomials used to generate the $s$-step bases. Additional bounds for Lanczos-based Krylov solvers running in finite precision can be found in~\cite{mest06}.

From these theoretical results on Krylov solvers running in finite precision, Simoncini/Szyld~\cite{Simoncini03} and Eshof/Sleijpen~\cite{Eshof2004} developed ``inexact Krylov subspace methods'' that apply the \spmv in lower precision to accelerate linear system solvers when this kernel dominates the cost of the computation. Theoretical results prove that inexact Krylov methods can achieve the same solution accuracy as high precision Krylov solvers, but little is known about the potential convergence delay.

Concerning long-recurrence strategies, Gratton et al~\cite{Gratton20} combined the previous findings from Bj\"{o}rck~\cite{bjorck67a} and Paige et al~\cite{Paige1980,Paige2006} to derive theoretical norms for a mixed precision GMRES algorithm based on modified Gram-Schmidt.
In this algorithm, they consider using inexact (e.g., single precision) inner products in the orthogonalization process, which results in a loss of double precision (DP-)orthogonality of the Krylov search directions. This makes the work by Gratton et al~\cite{Gratton20} very similar to our approach. However, our approach is different in several aspects:
\begin{itemize}
    \item We decouple the arithmetic precision from the memory storage format to maintain the orthogonal search directions in lower precision while preserving full precision in all computations;
    \item We consider not only IEEE single precision as the reference compact storage format, but also IEEE half precision (HP) and fixed point formats based on 32-bit and 16-bit integers;
    \item We realize a production-ready and sustainable implementation for high performance GPU architectures including restarting and classical Gram-Schmidt with reorthogonalization; and
    \item We provide comprehensive experimental results analyzing accuracy, convergence, and performance of our mixed precision GMRES solver.
\end{itemize}

%% file: s2-alg.tex
\section{The GMRES algorithm}
\label{sec:alg}

\begin{figure}
\begin{center}
\fbox{
\begin{minipage}[t]{0.9\textwidth} 
{\footnotesize
\begin{tabbing}
xxx\=xxx\=xxx\=xxx\=xxx\=\kill
\> \itshape  1. \' Compute $r_0 := b - Ax_0$, $\beta := \|r_0\|_2$, and $v := r_0 / \beta$. Set $V_1=[\,v\,]$\\
\> \itshape  2. \' \texttt{\bf for} $j := 1, 2, \ldots, m$ \\
\> \itshape  3. \' \> Compute $w := A (M^{-1}v)$ \\
\> \itshape  4. \' \> $\omega := \|w\|_2$ \\
\> \itshape  5. \' \> Orthogonalize $h_{1:j,j} := V_j^T w$, $w := w - V_j h_{1:j,j}$ \\
\> \itshape  6. \' \> $h_{j+1,j} := \|w\|_2$ \\
\> \itshape  7. \' \> \texttt{\bf if} $(h_{j+1,j} < \eta \, \omega)$ \texttt{\bf then}\\
\> \itshape  8. \' \> \> Re-orthogonalize $u := V_j^T w$, $w := w - V_j u$ \\
\> \itshape  9. \' \> \> $h_{1:j,j} := h_{1:j,j} + u$ \\
\> \itshape 10. \' \> \> $h_{j+1,j} := \|w\|_2$ \\
\> \itshape 11. \' \> \texttt{\bf endif} \\
\> \itshape 12. \' \> \texttt{\bf if} $(h_{j+1,j} = 0)$ \texttt{\bf or} 
                      $(h_{j+1,j} < \eta \, \omega)$ \texttt{\bf then} set $m := j$ and \texttt{\bf go to step 17}, \texttt{\bf endif} \\
\> \itshape 13. \' \> $v := w / h_{j+1,j}$ \\
\> \itshape 14. \' \> Set $V_{j+1} := \left[V_j,~v\right]$ \\
\> \itshape 15. \' \texttt{\bf endfor} \\
\> \itshape 16. \' Define the $(m + 1) \times m$ Hessenberg matrix $\bar{H}_m = \left(h_{ij}\right)_{1 \le i \le m+1, 1 \le j \le m}$ \\[0.05in]
\> \itshape 17. \' Compute $y_m$ the minimizer of $\|\beta e_1 - \bar{H}_m y\|_2$ and $x_m := x_0 + M^{-1} (V_m y_m)$ \\
\> \itshape 18. \' \texttt{\bf if} satisfied \texttt{\bf then Stop}, \texttt{\bf else} set $x_0 := x_m$ and \texttt{\bf go to step 1}, \texttt{\bf endif}
\end{tabbing}
}
\end{minipage}
}
\end{center}
\caption{Algorithmic formulation of the restarted GMRES algorithm for the solution of sparse linear systems.}
\label{fig:gmres}
\end{figure}

Consider the linear system 
\begin{equation}
Ax=b,
\label{eqn:linsys}
\end{equation}
where the coefficient matrix $A \in \Rnn$ is sparse, with $n_z$ nonzero entries;
$b \in \Rn$ represents the right-hand side vector; and $x \in \Rn$ contains the 
sought-after solution (vector). 
Figure~\ref{fig:gmres} displays a mathematical formulation of the restarted GMRES algorithm for the iterative solution of~\nref{eqn:linsys}.
There we assume that $M \in \Rnn$ defines an appropriate preconditioner;
$x_0$ is an initial approximation to the actual solution; 
$e_1$ stands for the first column of the square identity matrix of order $m+1$; and
the scalars $m$ and $\eta$ respectively define the dimension of the orthogonal
basis and the threshold for the re-orthogonalization.
The orthogonalization mechanism in the algorithm relies on the classical Gram-Schmidt (CGS) method, but a version that
employs the modified Gram-Schmidt (MGS) variant is simple to derive from that~\cite{GVL3}. We prefer CGS over MGS as it allows for higher efficiency (using BLAS 2 routines), and provides comparable accuracy if enhanced with optional re-orthogonalization.
The stopping criterion can be based, for example, on the residual
$\|r_m\|_2 = \| b-Ax_m \|_2$ being smaller than a certain relative threshold $\tau \cdot \|b\|_2$.
For convenience, the GMRES algorithm internally keeps track of the residual by iteratively updating the residual vector in every iteration. However, rounding effects can cause the iterative residual to differ from the explicit residual, and every restart therefore explicitly computes the residual to re-align the iteratively-computed residual.

From the computational point of view, the main kernels appearing in the GMRES algorithm 
correspond to the application of the preconditioner $M$ and the \spmv operation with the
coefficient matrix $A$ (both in Line~3);
the orthogonalization of vector $w$ with respect to the vectors in the basis $V_j$
(Lines~5 and~8); the solution of the linear least squares (LLS) problem (Line~17); the assembly of the
next iterate, which requires the application of the orthogonal basis followed by the preconditioner (Line~17);
and a few minor vector operations such as \textsc{axpy}s, vector scaling, etc.~\cite{blas1}.

The LLS problem in the GMRES algorithm can be solved via the QR factorization~\cite{GVL3},
where this decomposition can be cheaply obtained using an updating technique as the Hessenberg matrices for two consecutive 
iterations basically differ only in one column.
Therefore, the cost associated with the solution of this problem is minor in comparison with that of the global
algorithm. 
In addition, the operations that are necessary to update the new estimate to the solution $x_m$ (Line~17) also
contribute a minor cost to the overall procedure, as they are $m$ times less frequent in comparison with
the kernel calls in Lines~3, 5, and~8.

%% file: s3-newalg.tex
\section{CB-GMRES storing the orthonormal basis in reduced precision}
\label{sec:newalg}
For simplicity, 
consider that the GMRES algorithm integrates
a simple preconditioner, such as a Jacobi scheme (or a block-Jacobi variant with a small block size)~\cite{Saa03}.
The performance of the algorithm is then strongly determined by the costs of
the \spmv kernel and the general matrix-vector products (\gemv), with $V_j^T$ and $V_j$.
These are memory-bound kernels, with their execution times largely dictated by the number of memory accesses (memory operations, 
or memops hereafter). 
The optimization we propose thus 
aims to reduce the cost of the \gemv operations by storing the orthogonal
basis $V_j$ in a more compact, reduced-precision format.

In order to analyze the theoretical memop count of the \spmv and the \gemv kernels, 
for simplicity, let us assume the following:
\begin{enumerate}
\item The right-hand side vectors for both types of matrix-vector products reside in cache. In general, this is not
true but, for the following theoretical derivation, the memory layer where the vectors reside is not important.
\item The sparse coefficient matrix is stored in the compressed sparse row (CSR) format.
      This is a general and flexible data layout that employs one integer per nonzero value to represent its column index, plus $n+1$ integers for the row pointers~\cite{Saa03}.
\item The re-orthogonalization mechanism included in the GMRES algorithm
      (Li\-nes 7--11 in Figure~\ref{fig:gmres}) is not needed. 
\end{enumerate}
Then, the ratio between the contributions of \spmv and the two \gemv to the memop count,
due to the accesses to the corresponding matrices,
is given by
\begin{equation}
{\footnotesize
\frac{\textrm{Memops \gemv}}{\textrm{Memops \spmv}} = \frac{2nm'}{n_z (1+f) + (n+1) f} \approx
\frac{2nm'}{ns (1+f) + nf}  = \frac{2m'}{s(1+f)+f},
}
\label{eq:memops}
\end{equation}
where 
$s=n_z/n$ is the average number of nonzero entries per row of the sparse matrix;
$m'=j-1$ is the size of the already-computed Krylov subspace, that is, the number of vectors the new search direction is orthogonalized against; and
$f>1$ represents a factor for the indexing overhead into the sparse data structures.
(For example,  when using 32-bit integers to represent the indices and 64-bit for the data values, $f=32/64 = 1/2$.)

For a non-restarted version of GMRES, the size of the Krylov supspace $m'$ steadily grows with the iteration count ($m'=j-1$ at iteration $j$), which hints that the memops related to the orthogonalization can quickly to dominate the cost. In practical implementations though, the GMRES solver is usually enhanced with a restart mechanism like in the formulation of the algorithm in Section~\ref{sec:alg}, to keep both the memory requirements and the orthogonalization cost at reasonable levels. Depending on the problem size and the available resources, the typical values for the restart parameter vary between $m=$ 30 and 200. At the same time, the nonzero-per-row ratio $s$ is usually relatively small, and often significantly smaller than the restart parameter $m$. 
Therefore, 
assuming a restart parameter $m$ and considering the memops in that restart cycle, equation~\eqref{eq:memops} then becomes
\begin{equation}
\label{eq:memops2}
{\footnotesize
\frac{\textrm{Memops \gemv}}{\textrm{Memops \spmv}} = \frac{\sum_{j=1}^{m-1}2nj}{m\left(n_z (1+f) + (n+1) f\right)} \approx
\frac{m}{s(1+f)+f}.
}
\end{equation}
With typical parameters $f=1/2$ and $m=100$, the memory access count due to the orthogonalization theoretically thus exceeds the memory access overhead for the \spmv kernel for matrices with ratios $s=n_z/n>67$.\\

\noindent\textbf{CB-GMRES.}
In order to reduce the memory access volume in the orthogonalization step of GMRES, we propose to store the vectors of the orthogonal basis $V_j$ in a compact reduced-precision format; retrieve the data from memory in that format; and transform the values into \textsc{ieee} 64-bit double precision (DP) prior to the
orthogonalization computations they are involved in (Lines 5, 8, and 17). This adheres to the idea of decoupling the memory storage format from the arithmetic precision, while preserving \textsc{ieee} 64-bit precision in the arithmetic operations~\cite{ijhpca-decoupling}.

The decoupling strategy provides full flexibility in terms of choosing a memory representation format, enabling the usage of 
the natural \textsc{ieee} 16-bit or 32-bit formats as well as other, non-standard  alternatives 
(with no hardware support for the arithmetic).
In particular, the property that the entries of the orthonormal vectors forming the Krylov basis are all bounded by 1 pushed us to the explore the efficiency of more aggressive customized formats. 
For example, it is possible to reduce the number of bits employed for the exponent in the floating-point format by normalizing them with respect to a baseline factor. 
In our investigation, we take this approach to the extreme,
resulting in the evaluation of fixed-point formats for the storage of the orthogonal basis.
For this purpose:
1) we normalize each vector of the basis by scaling its entries with (the inverse of) its largest vector entry (in absolute value); and 2) 
we then store only the fractional part of each value of the result
as an integer number, plus the normalization factor for each vector.

For convenience, we refer to the resulting algorithm as ``compressed basis GMRES (CB-GMRES)'' in the remainder of the paper even though we emphasize, that we still use DP in all arithmetic operations and only use lower precision formats for the memory operations.

\noindent
\textbf{Discussion.}
Storing the orthogonal basis of a Krylov method in a reduced precision format will typically introduce rounding errors that may affect the numerical properties of the method, potentially impacting the convergence and numerical stability of the iterative solver.
As the solution approximation is optimal in the generated Krylov subspace, 
perturbed Krylov search directions may result a loss in the DP-orthogonality of the search directions and a different (Krylov) subspace and in which the solution approximation is computed. However, the solution approximation process accounts for the perturbed search directions, and as long as the generated subspace allows for a good approximation of the solution, this approximation will be found in the optimization process. Hence, as long as the search directions are ``relatively'' close to the optimal search directions, 
the convergence will only be mildly affected. In particular, we may assume that the need for additional search directions (equivalent to additional iterations) can be compensated by the faster execution of each iteration.

To close this section, we emphasize that:
\begin{list}{--}{}
\itemsep=0pt\parskip=0pt
\item Our approach is orthogonal and complementary to
other techniques which aim to reduce the memory access overhead, for example, by
customizing the sparse matrix data layout to the application,
operating with iterative refinement scheme+mixed precision for the coefficient matrix, or
the exploiting customized precision in the preconditioner, among others. 
\item The arithmetic precision is decoupled from the representation format so that we can actually store the data for the orthogonal basis in any format while relying on the data types with hardware support for the arithmetic operations.
\end{list}

%% file: s4-impl.tex
\section{Implementation of CB-GMRES}
\label{sec:implementation}
\subsection{The Ginkgo sparse linear algebra library}
\label{ginkgo}
\begin{figure}[!t]
  \centering
  \includegraphics[width=0.9\linewidth]{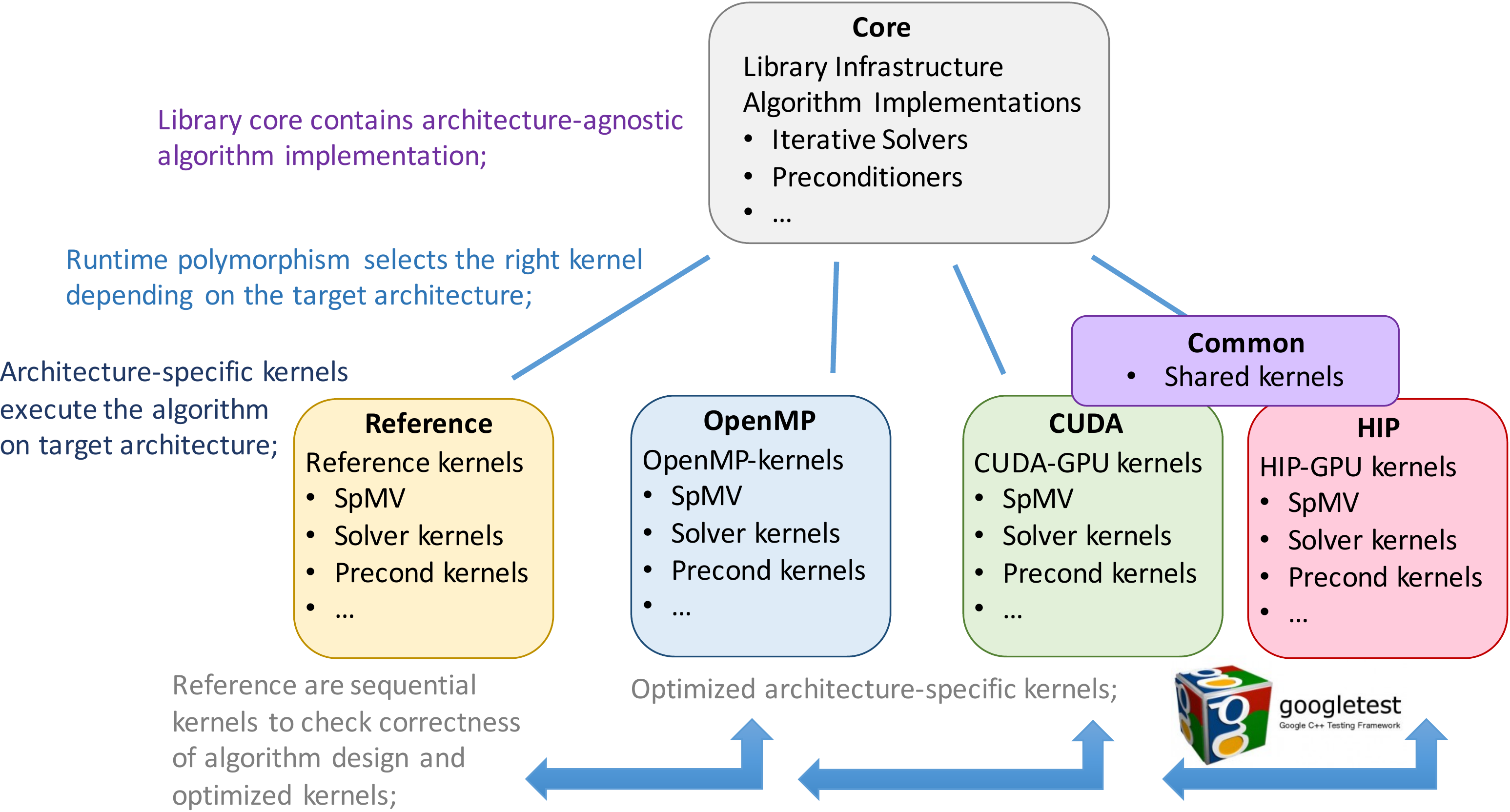}
  \caption{The architecture of the \gko~library separating the algorithmic core from the backends.}
  \label{fig:gko_design}
\end{figure}

For convenience and ease of use, we have realized the CB-GMRES algorithm in the \gko~ecosystem.
\gko~is a sparse linear algebra library implemented in modern C++ 
that embraces two principal design concepts~\cite{anzt2020ginkgo}:
The first principle, aiming at future technology readiness, is to consequently 
separate the numerical algorithms from the hardware-specific kernel 
implementation to ensure correctness (via comparison with sequential 
reference kernels), performance portability (by applying hardware-specific 
kernel optimizations), and extensibility (via kernel backends for other 
hardware architectures); see Figure~\ref{fig:gko_design}. 
The second design principle -- pursuing user-friendliness -- 
is the convention to express functionality in terms of linear 
operators: every solver, preconditioner, factorization, matrix-vector product, and matrix reordering is expressed as a linear operator (or composition thereof). 
This allows to easily combine the CB-GMRES with any preconditioner available in \gko, and to select the realization of the \spmv kernel that is most appropriate for the characteristics of a specific problem~\cite{topcspmv}. 

\gko~relies on an ``executor'' concept to favor platform portability. 
The executor specifies the memory location and execution domain of linear algebra objects and abstracts the computational capabilities of distinct devices. 
Each executor implements methods for allocating/deallocating memory on
the device targeted by the executor, copying data between executors, providing hardware-specific kernels, running operations, and synchronizing all operations launched on the executor.
The user can run a single code on different platforms (without having to modify his/her code) 
by selecting the proper executor at the beginning of the application. This
encapsulates all information in the executor, and automatically orchestrates memory allocation, memory transfers, and kernel selection. 
For the CB-GMRES implementation with the orthonormal Krylov basis stored in reduced precision, the executor concept is extended with a ``memory accessor'', described next, that handles the data conversion transparently to the user.

\begin{figure}
    \centering
    \includegraphics[width=.6\columnwidth]{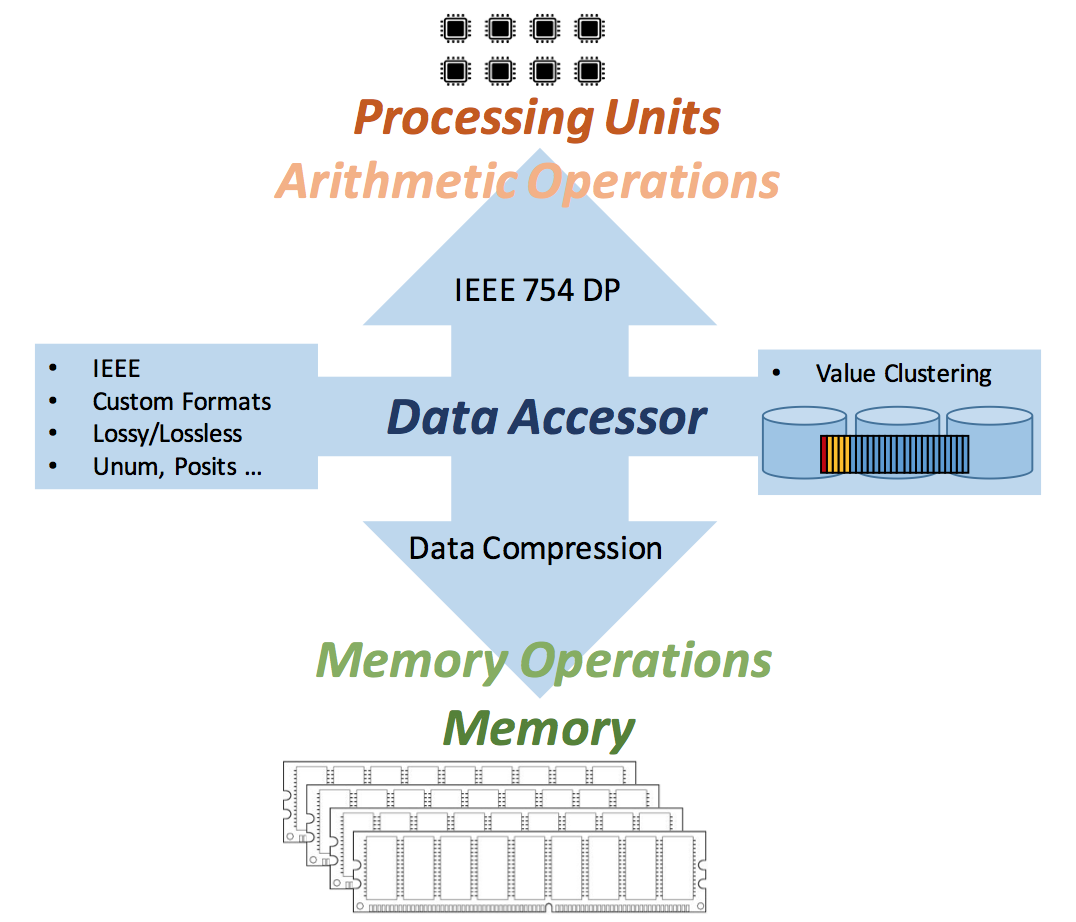}
    \caption{Accessor separating the memory format from the arithmetic format and realizing on-the-fly data conversion in each memory access.}
    \label{fig:accessor}
\end{figure}

\noindent
\subsection{Memory accessor}
At a high level, the idea of the CB-GMRES solver is to compress the orthogonal matrix/vector before and after the memory operations using one of
the reduced/customized storage formats, but still  use the working precision  (i.e., DP) for the arithmetic operations.
Retrieving the orthonormal basis in reduced precision from memory thus requires reading the basis contents and converting them into DP. 
When these values are stored in SP, the conversion is easy to perform via a datatype casting operator. For
fixed point representations, though, the conversion requires some additional manipulations plus the scaling with a normalization factor.

To decouple the memory access and conversion from the code development effort, we use a memory accessor that converts the data
between DP and the memory storage/communication format on-the-fly (see Figure~\ref{fig:accessor}).
The efficient implementation of the accessor aims to hide the cost of these data conversions by overlapping them with the memory accesses, in principle introducing a minor or even negligible overhead. In addition, the introduction of this technique can accelerate the execution as accessing the data in lower precision significantly reduces the memory access volume per iteration.

Considering the realization of the CB-GMRES algorithm,
after the new basis vector $v_j=v$ is formed at iteration $j$, the memory accessor is activated in order to compress the DP values of this vector before storing them into memory; see Lines~13 and~14 in the algorithm in Figure~\ref{fig:gmres}. The memory accessor is also active when retrieving
the contents of the full orthogonal basis $V_j$ from memory; see Lines~5 and~8 of the algorithm.

On the technical side, the accessor leverages static polymorphism (via C++ templates) for both the arithmetic precision (in our work, fixed to IEEE DP) and the memory format. While this flexible design can accommodate any memory format, we currently only support \texttt{<float64>}, \texttt{<float32>} and \texttt{<float16>} (for IEEE DP, SP and HP, respectively), and \texttt{<int32>}, \texttt{<int16>} (for 32-bit and 16-bit fixed point formats) in \gko. The versions based on integer formats rely on a fixed point representation in order to maintain the orthonormal basis vectors. This representation only requires a fractional part because the vectors are normalized, making each vector entry smaller than 1. However, this is not efficient for large vectors because the largest absolute value will likely be significantly smaller than 1, therewith wasting representation range (and precision). To optimize the accuracy, a different scaling factor is used for each vector:
\[\sigma_j = \frac{\|v_j\|_\infty / \|v_j\|_2}{\texttt{max\_intxx}},\]
where $v_j$ is the vector computed at iteration $j$ before normalization, and \texttt{max\_intxx} is the maximum positive value of the integer representation using \texttt{xx} $\in\{16,32\}$ bits. Both norms can be computed simultaneously so that the extra overhead due to the memory accesses to obtain the infinity norm remains small. The vector $v_j$ is then stored in $V_{j+1}$ as
\[V_{j+1} = \left[V_{j}, v_j / \sigma_j \right], \]
and any subsequent access to the contents of $V_{j+1}$ implies an intrinsic post\hyph multiplication by a diagonal matrix $\Sigma_{j+1} = \text{diag}(\sigma_1,\sigma_2,\ldots,\sigma_j,\sigma_{j+1})$ that
contains the scaling factors on the diagonal.
This scaling adds one multiplication per element to the computational cost of any operation involving the orthogonal basis, and storing the scaling factor in memory. However, as the whole algorithm is heavily bandwidth bound, we expect the overhead remaining small.

%% file: s5-experiments.tex
\begin{table}
\caption{Test matrices}
\label{tab:matrices}
\centering
\resizebox*{!}{0.95\textheight}{%
\begin{tabular}{lrrr}
\toprule \bfseries Matrix &
\multicolumn{1}{c}{\bfseries Size} &
\multicolumn{1}{c}{\bfseries Non-zeros} &
\multicolumn{1}{c}{\bfseries Non-zeros per row} \\
\midrule
\input{tables/nnz}
\bottomrule
\end{tabular}}
\end{table}

\section{Experimental Evaluation of the compressed basis GMRES}
\label{sec:experiments}
In this section, we analyze several properties of the CB-GMRES algorithm in order to assess the benefits of this solver as part of production code. Concretely, we investigate the following questions:
1) Can we achieve high accuracy in the solution approximations? 2) How significant is the convergence delay introduced by moving away from the ``full'' precision Krylov search directions and utilizing instead search directions that are low precision approximations of these orthonormal vectors? 3) What are the performance advantages of the CB-GMRES over the standard (DP) GMRES? 4) Which specific storage format we should use for the memory operations?

\subsection{Setup}

To answer these questions, we select a set of 49 large-scale test matrices from the Suite Sparse Matrix Collection~\cite{suitesparse} that we adopt as benchmark problems to explore the accuracy, convergence, and performance of the CB realizations of the GMRES algorithm. The selected test matrices are regular, appropriate in size and nonzero count, and a DP GMRES needs at least 40 iterations to converge. The test matrices are listed along with some key properties in Table~\ref{tab:matrices}. 

The CB-GMRES algorithm is implemented utilizing building blocks from the \gko~environment. The orthogonalization kernel is based on classical Gram-Sch\-midt with optional re-orthogonalization. All other functionality (\spmv kernels, preconditioners, utility functions, comparison solvers, etc.) is taken from the \gko~library. Unless otherwise stated, we enhance all the CB and DP GMRES algorithms with a simple light-weight scalar Jacobi preconditioner (diagonal scaling) as this generally improves convergence and provides a more realistic setting than a stand-alone GMRES algorithm. The \spmv kernel integrated in all variants of GMRES to generate the Krylov search directions is \gko's CSR-based \spmv routine; this particular realization of \spmv maintains the coefficient matrix in Compressed Sparse Row (CSR) format, and automatically selects a \csr kernel that is optimal for a problem-specific sparsity pattern~\cite{spmvtopc}.
The DP GMRES code is identical to the CB-GMRES code with the orthogonal basis stored in DP 
as we did not detect any runtime overhead from using the memory accessor.

In the performance tests, we utilize \gko's CUDA executor, which is heavily-optimized for NVIDIA GPUs. We run all experiments on an NVIDIA V100 GPU with support for compute capability 7.0~\cite{volta}. The V100 accelerator board is equipped with 16 GB of main memory, 128 KB L1 cache and 6MB of L2 cache. Bandwidth tests achieved 897 GB/s for main memory access in this particular device~\cite{spmv_isc20}. 
The theoretical peak performance for the V100 GPU is 7.83 DP TFLOPS (that is, 
$7.83\cdot 10^{12}$ floating-point operations per second). 
\gko's CUDA backend was compiled using CUDA version~9.2.

\subsection{Accuracy of CB-GMRES}
We initially investigate whether CB-GMRES can match the accuracy levels attained by DP GMRES. For that purpose, we consider 49 linear systems of the form $Ax=b$, with the coefficient matrix defined from the test matrices in Table~\ref{tab:matrices}, and the components of the right-hand side vector set as $b_i=sin(i)$. The GMRES algorithm is started with an initial guess $x_0=0$, uses a restart parameter $m=100$, and is considered to converge when the solution approximation $x^*$ yields a residual $\|Ax^*-b\|_2 \leq 10^{-9}\|b\|_2$. 
We believe this setting reflects real-world problems, and we use it for the rest of the evaluation.

To avoid expensive explicit residual computations, the GMRES algorithm internally updates a recurrence residual that is used to check convergence. However, when using finite precision and due to the accumulation of rounding error, this iteratively-computed residual can diverge from the real residual, and the GMRES algorithm may stop ``too early'' even though the real residual did not fall below the selected threshold. Using the compressed basis formats to store the orthonormal basis may enhance this effect. To tackle this problem, we modify all the implementations to compute the explicit residual once convergence is indicated by the recurrence residual, but continue iterating with the updated residual in case the actual accuracy threshold is not fulfilled.

\begin{figure}
  \centering
  \includegraphics[width=\columnwidth]{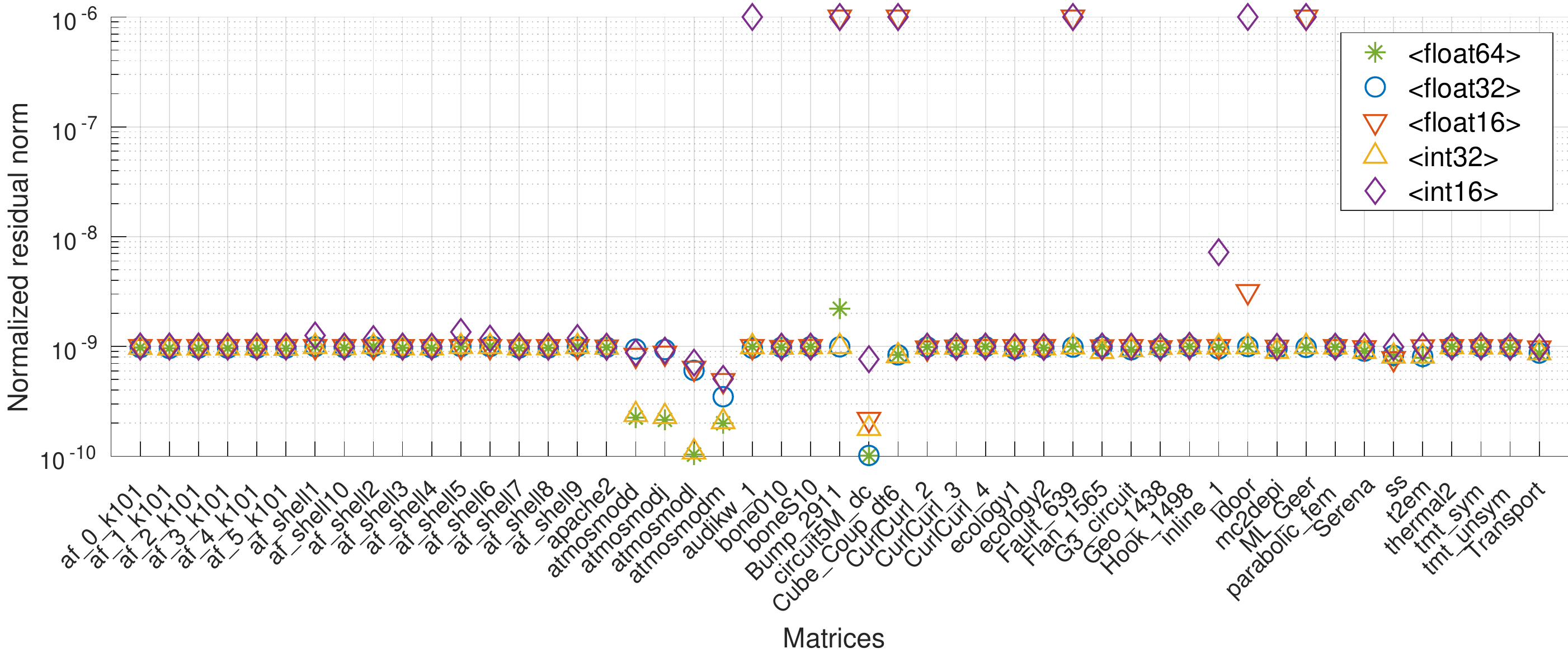}
  \caption{Normalized residual of the distinct CB-GMRES versions.} 
  \label{fig:accuracy}
\end{figure}

To assess the solution accuracy, in Figure~\ref{fig:accuracy} we report the normalized residual 
$\|Ax^*-b\|_2/\|b\|_2$ for the solution approximations 
computed with the distinct CB-GMRES versions. 
In all figures and tables in this section, as well as in the following discussions, 
\texttt{<floatxx>} and
\texttt{<intxx>} respectively identify different realizations of CB-GMRES with the orthogononal basis stored using
\texttt{xx}-bit floating-point and fixed-point formats.
The notation \texttt{<float64>} refers to the DP GMRES algorithm.
In these initial results, 
we notice that the CB-GMRES variants fulfill the residual accuracy requirement in most cases, but small differences in the residual norms may indicate variations in the convergence rate.

\subsection{Convergence of CB-GMRES}
\begin{figure}
  \centering
  \begin{tabular}{lr}
    \includegraphics[width=.46\columnwidth]{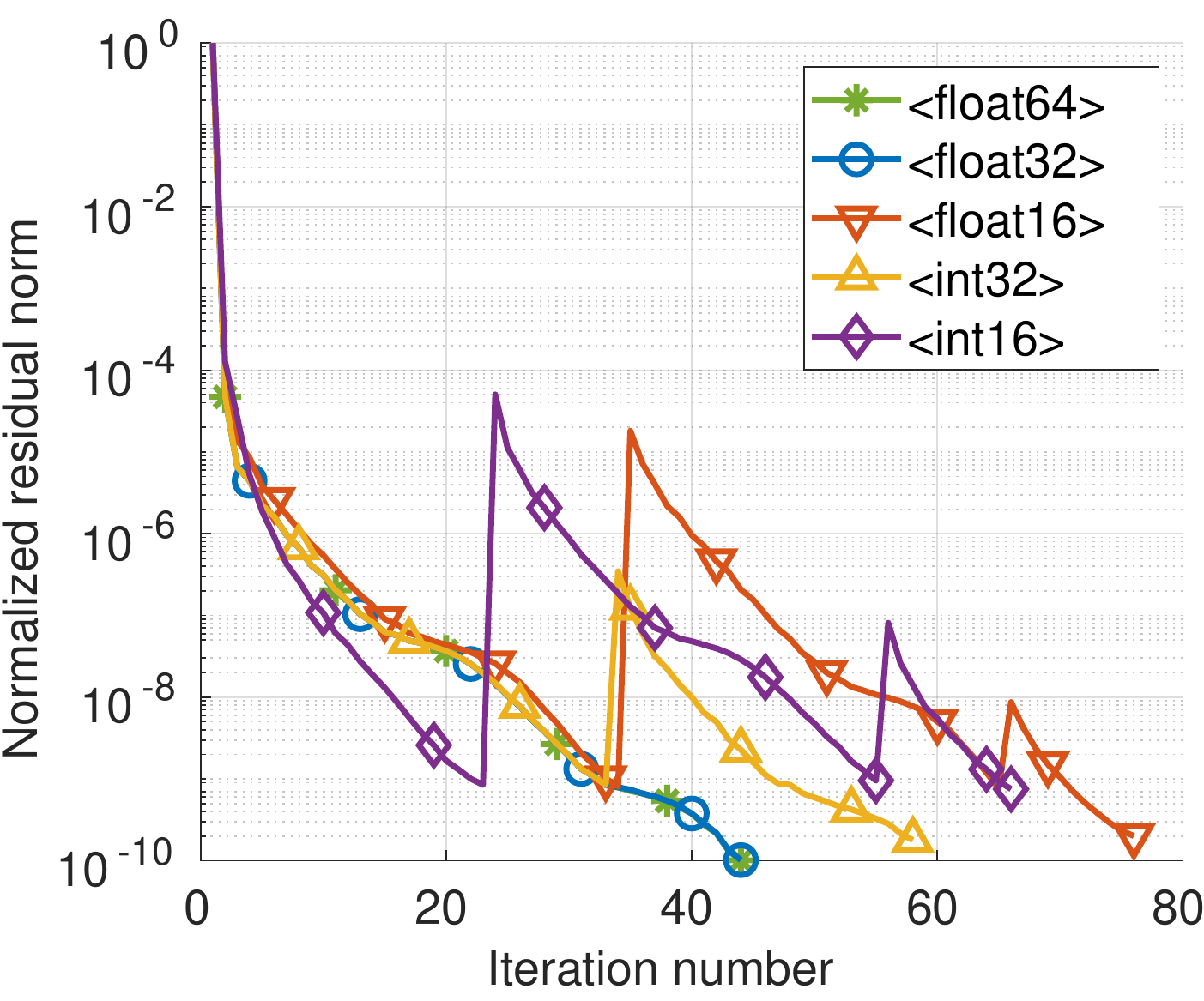}
    &
    \includegraphics[width=.46\columnwidth]{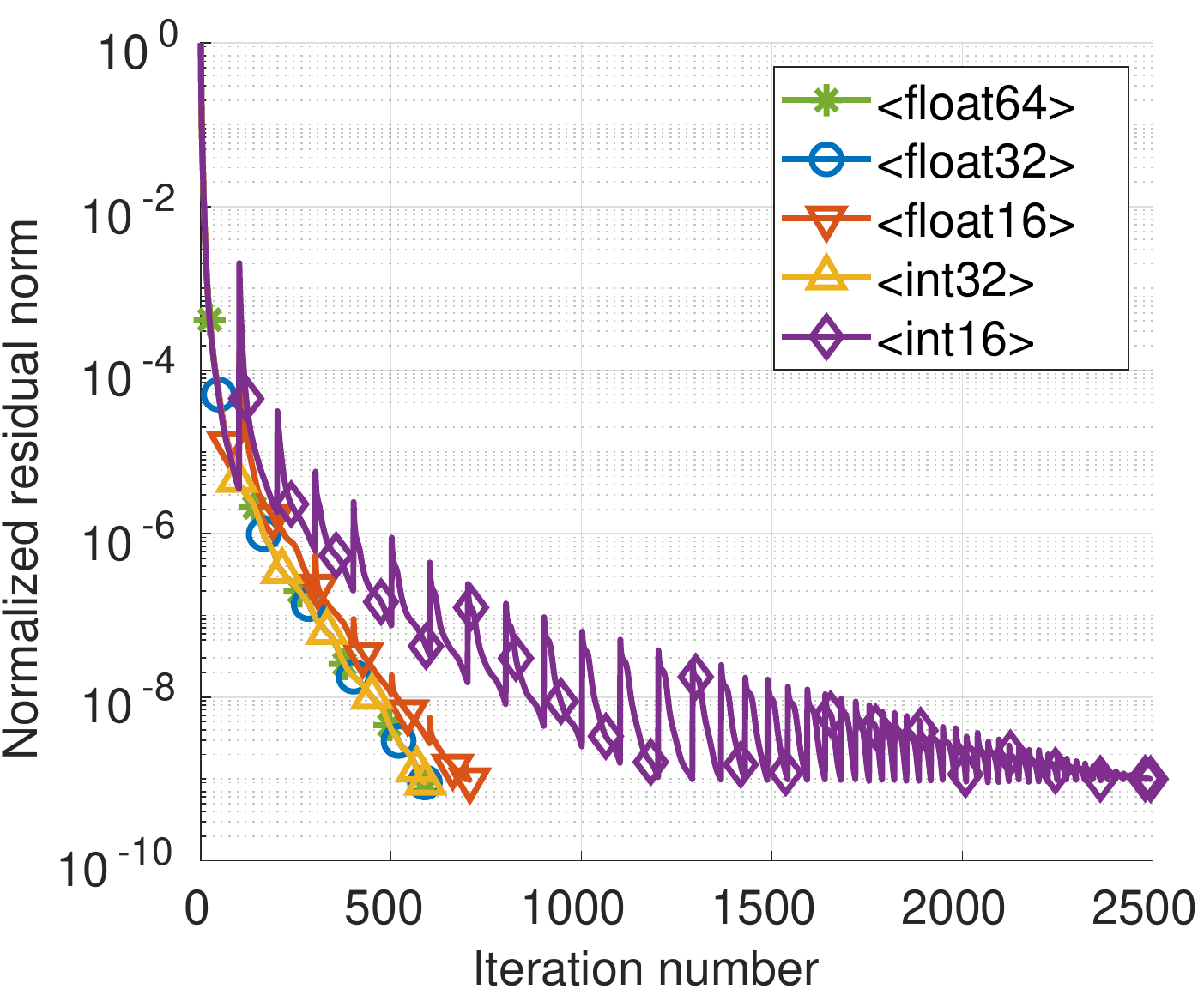}
  \end{tabular}
  \caption{Convergence of the CB-GMRES variants for the \texttt{circuit5M\_dc} and \texttt{Serena} problems.}
  \label{fig:convergence}
\end{figure}

\begin{figure}
  \centering
  \includegraphics[width=\columnwidth]{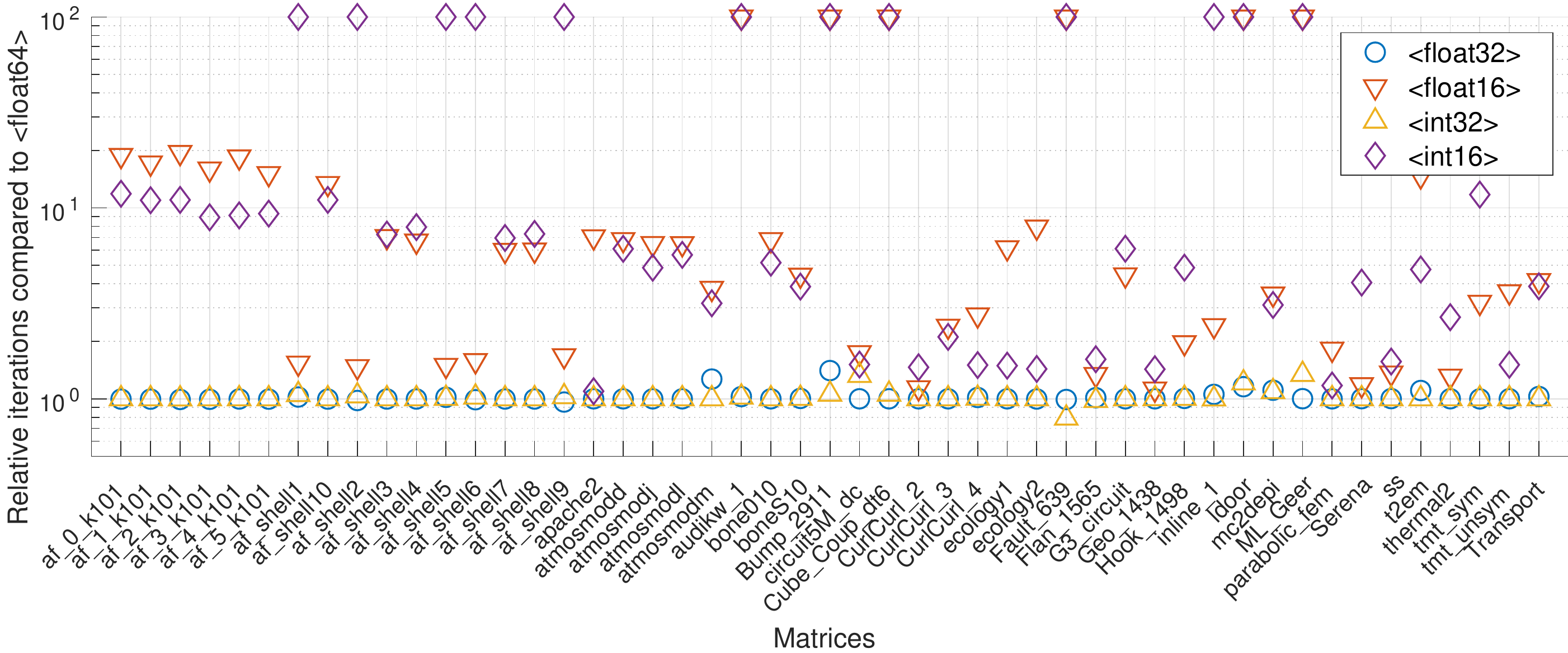}
  \caption{Iteration overhead of the CB-GMRES variants relative to the DP GMRES iteration count for a residual threshold $\|Ax^*-b\|_2\leq 10^{-9}\|b\|_2$.}
  \label{fig:iterations}
\end{figure}

In Figure~\ref{fig:convergence} we expose the convergence behaviour of the CB-GMRES variants for the \texttt{circuit5M\_dc} and \texttt{Serena} problems. (Similar behaviour was observed for other problems from the 49-case collection.) While in this case all CB-GMRES variants achieve the same final accuracy, the storage format selected for the orthogonal basis impacts the convergence rate and, in consequence, the iteration count. In addition, the spikes in the residual curves identify the restart points that update the recurrence residual with an explicitly computed residual. For \texttt{GMRES<int16>} in particular, this results in significant corrections of the normalized residual.
As expected, using a compressed format to store the orthogonal basis can delay convergence and require additional search directions. 
In order to quantify this effect, in Figure~\ref{fig:iterations} we display the iteration count of the CB-GMRES variants relative to the DP GMRES iteration count. An iteration overhead of 100 in that figure identifies those storage formats for which CB-GMRES did not converge within the iteration limit.

This experiment shows that the realizations
\texttt{GMRES<float32>} and \texttt{GMRES<int32>} match the iteration count of DP GMRES in almost all cases, and only need a few additional iterations for a couple of problems. In contrast, when the orthogonal basis is stored using the 16-bit formats, the overhead often increases dramatically, and even for matrices within the same (\texttt{af\_shell}) group, there is no clear winner between the \texttt{GMRES<float16>} and \texttt{GMRES<int16>}. 
As expected, for those problems where \texttt{GMRES<float32>} and \texttt{GMRES<int32>} need additional iterations, \texttt{GMRES<float16>} and \texttt{GMRES<int16>} typically fail.

\begin{figure}
  \centering
  \begin{tabular}{lr}
    \includegraphics[width=.46\columnwidth]{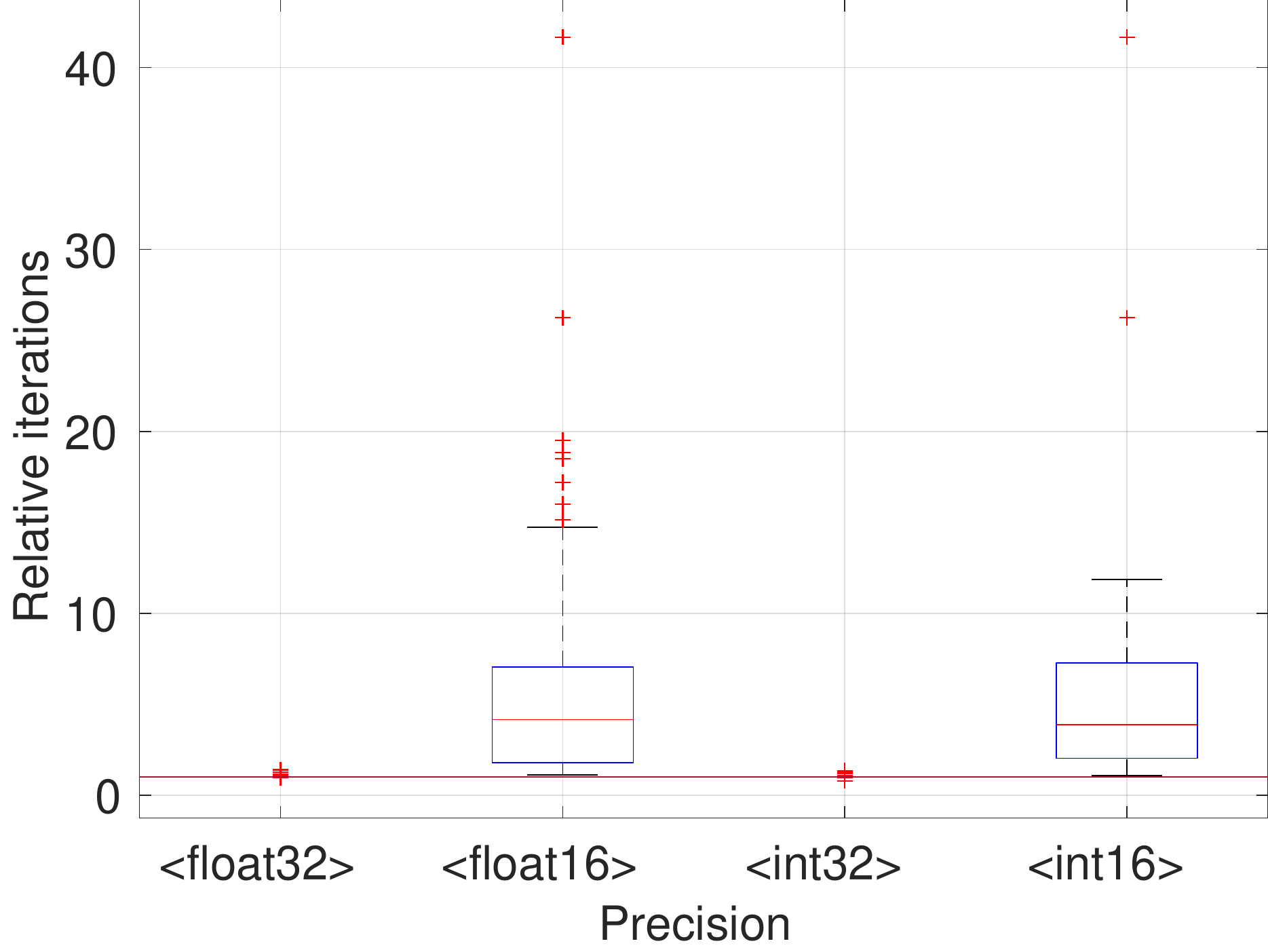}
    &
    \includegraphics[width=.46\columnwidth]{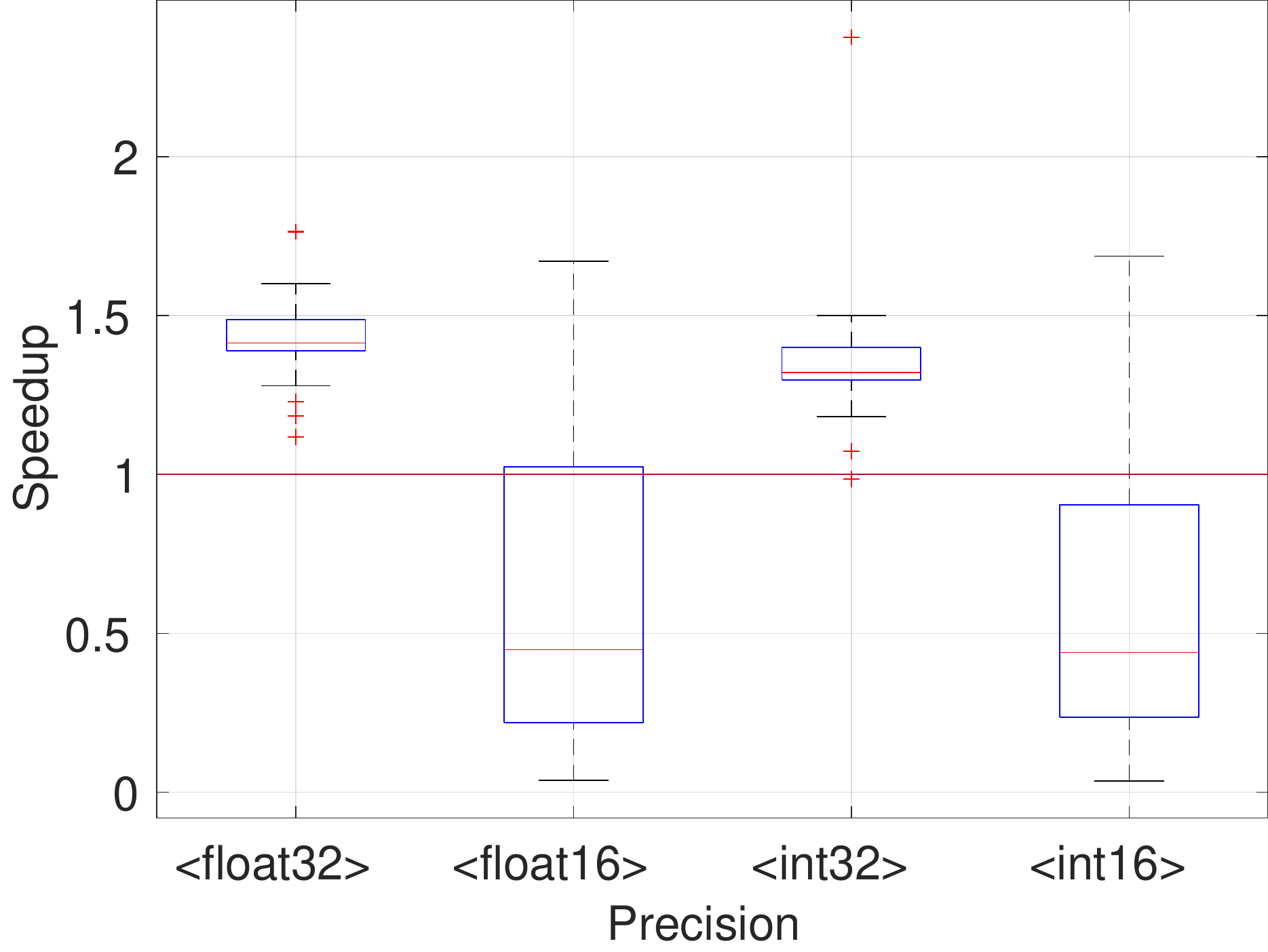}
  \end{tabular}
  \caption{Statistics obtained from running the CB-GMRES algorithms on the 49 test problems. Left: Iteration overhead (relative to DP GMRES); Right: speedup relative to DP GMRES.}
  \label{fig:statistics}
\end{figure}

In the left-hand side plot in Figure~\ref{fig:statistics} and Table~\ref{tab:statisticstime} (left-hand side), we report a few key statistics obtained from the experimental evaluation with the 49 test problems. While storing the vector entries in \texttt{<float32>} or \texttt{<int32>} incurs no iteration overhead, when using 16-bit storage we obtain a median iteration overhead of 4$\times$, with the 50\%-quantiles varying between 2$\times$ and $7.5\times$, and the 90\%-quantiles reaching up to 12$\times$ and 15$\times$ for \texttt{<int16>} and \texttt{<float16>}, respectively.

\begin{table}[]
    \centering
    \footnotesize
    \begin{tabular}{l|r|r|r}
        Solver & arithmetic mean & arithmetic median & variance \\
        \hline
        \texttt{GMRES<float64>} & 1    & 1    &  0 \\ 
        \texttt{GMRES<float32>} & 1.02 & 1    &  0.01 \\
        \texttt{GMRES<float16>} & 6.97 & 4.16 & 62.30 \\
        \texttt{GMRES<int32>} & 1.02   & 1    &  0.01 \\
        \texttt{GMRES<int16>} & 5.86   & 3.88 & 46.95 \\
    \end{tabular}
    \caption{Statistics for the \texttt{GMRES<storage\_format>} iteration count normalized to the \texttt{GMRES<float<64>>} implementation on the test matrices listed in Table~\ref{tab:matrices}.}
    \label{tab:statisticstime}
\end{table}

\subsection{Performance of CB-GMRES}
Even though we now have experimentally demonstrated that the CB-GMRES variants can compensate for the perturbations in the subspace via additional iterations (which is equivalent to extending the subspace by additional search directions), the resulting algorithms will only be useful in production if the associated iteration overhead is smaller than the runtime reduction coming from the decreased memory access volume. In the right-hand side plot in Figure~\ref{fig:statistics} we show statistics on the performance improvements that CB-GMRES renders over DP GMRES when using different storage formats for the orthogonal basis. 
As could be expected from the large iteration overheads, storing the orthogonal basis in \texttt{<int16>} or \texttt{<float16>} usually results in a slowdown of the global solution process. Conversely, storing the orthogonal basis in \texttt{<int32>} or \texttt{<float32>} yields attractive performance improvements, with slight advantages for the \texttt{GMRES<float32>} variant. The median speedup for \texttt{GMRES<float32>} is 1.4$\times$, with the 50\%-quantiles reaching up to 1.6$\times$ and outliers reaching up to 1.75$\times$. Here we note that \texttt{GMRES<int32>} shows an outlier with a 2.4$\times$ speedup, which is likely related to faster convergence due to ``lucky rounding.''

\begin{figure}
  \centering
  \includegraphics[width=\columnwidth]{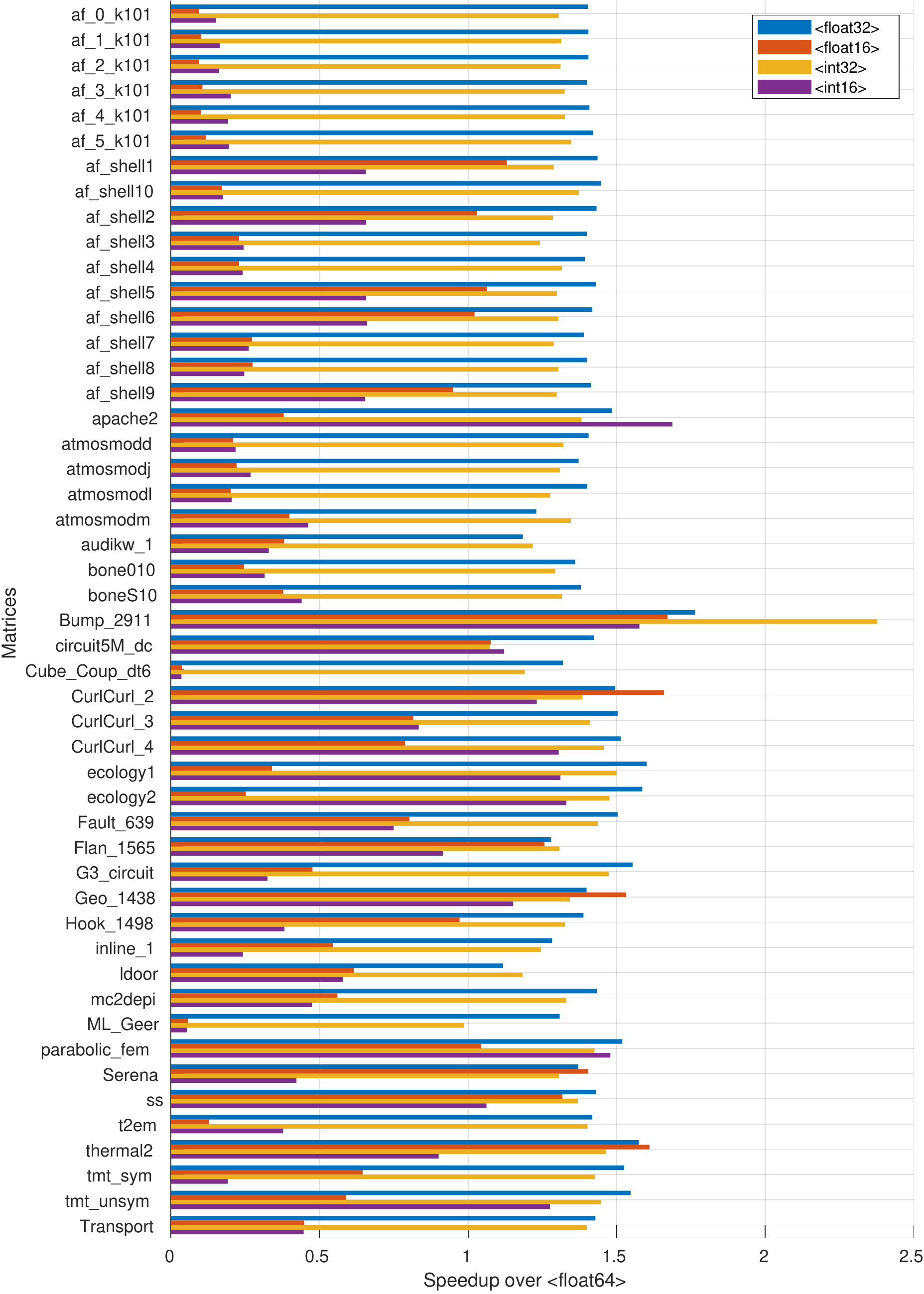}
  \caption{speedup of the CB-GMRES variants over DP GMRES for a residual threshold $\|Ax^*-b\|_2/ \leq 10^{-9} \|b\|_2$.}
  \label{fig:speedup}
\end{figure}

In Figure~\ref{fig:speedup} we provide a detailed performance evaluation by visualizing the speedup for the distinct test problems. There we notice a very uniform picture concerning the speedups for \texttt{GMRES<float32>} and \texttt{GMRES<int32>}, with \texttt{GMRES<float32>} being slightly superior. This is likely due to the overhead of the scaling process and the additional scaling factors needed when storing the basis vectors in \texttt{GMRES<int32>}. From this experiment, we conclude that the \texttt{GMRES<float32>} is an appropriate choice for a wide range of problems.

\begin{figure}
  \centering
  \includegraphics[width=\columnwidth]{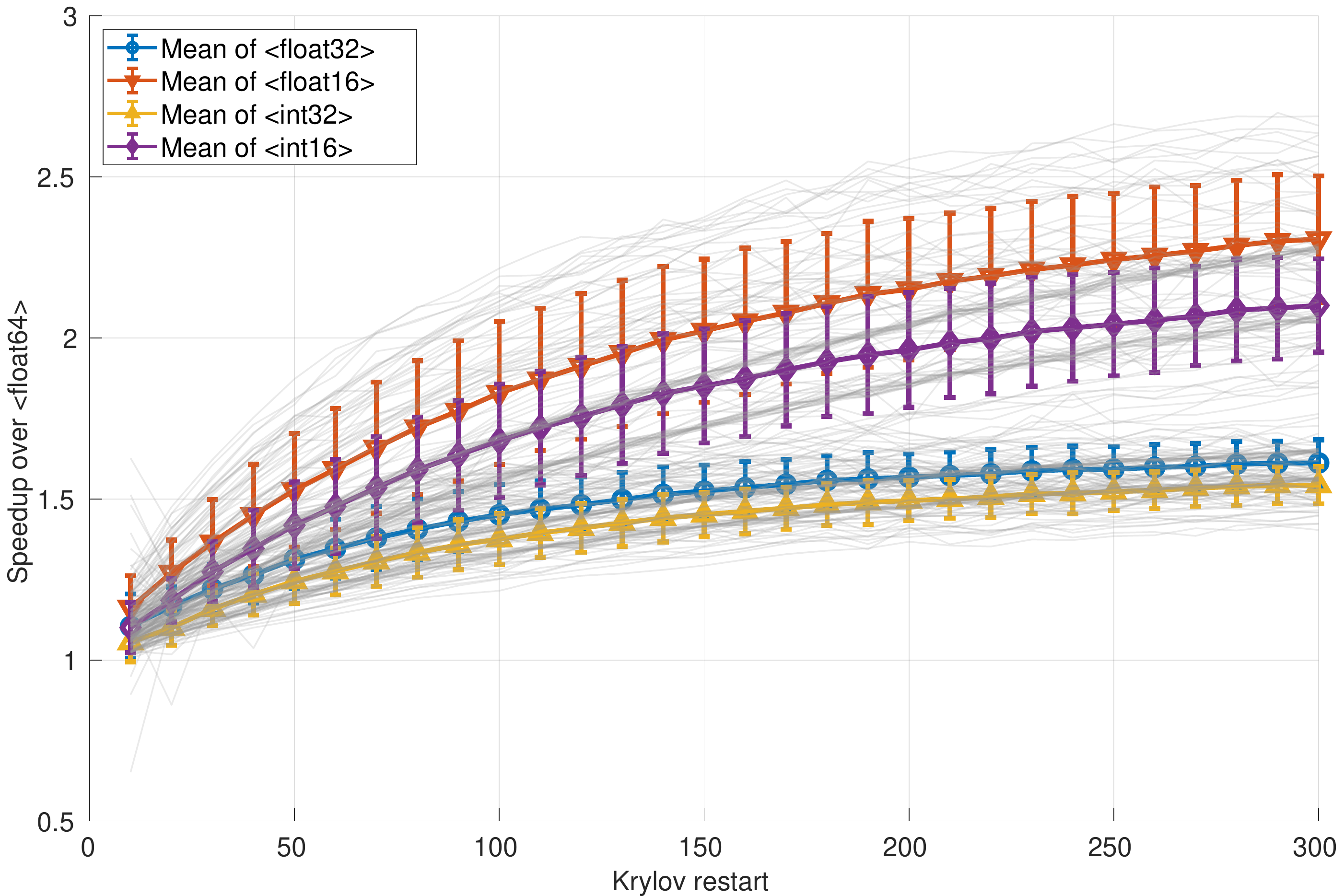}
  \caption{speedup for different CB-GMRES variants (\texttt{GMRES<precision\_format>}) over DP GMRES (\texttt{GMRES<float64>}) for increasing restart values.}
  \label{fig:restart}
\end{figure}

When motivating the use of a more compact storage format to maintain the orthonormal vectors in Section~\ref{sec:newalg}, we argued that the memory savings against DP GMRES grow with the size of the Krylov subspace; that is, the instances of CB-GMRES using a larger restart parameter $m$ should attain larger performance benefits than their CB-GMRES counterparts adopting smaller restart values.
In more detail, when ignoring numerical effects, we can expect that the speedup asymptotically reaches the ratio between the storage format complexities: $4\times$ when using \texttt{GMRES<float16>} or \texttt{GMRES<int16>}; and $2\times$ when using \texttt{GMRES<float32>} or \texttt{GMRES<int32>}. In Figure~\ref{fig:restart} we quantify those speedups experimentally, considering restart parameters in the range 10--300. We note that restart values beyond 200 are rarely employed as they introduce numerical instabilities and significant memory- and computational overhead. To avoid this issue, this experiment considers the runtimes needed to execute 10 restart cycles but ignores any numerical effects. Also, even though we already identified the \texttt{GMRES<float32>} as being superior in terms of convergence and performance, we include all CB variants in this analysis. 
In Figure~\ref{fig:restart} we employ grey lines to indicate the speedup behavior for the distinct matrices and use boxplots to illustrate the statistics for the CB-GMRES variants.
The results indicate that the average speedups for \texttt{GMRES<float32>} or \texttt{GMRES<int32>} asymptotically approach a value below $2\times$, with the speedups being constantly higher for the former (which requires no scaling). The speedup is smaller than 2$\times$ because the cost savings are limited to those obtained from the compressed storage of the orthogonal basis, but not in other parts of the algorithm such as, for example, the \spmv kernel (see Amdahl's law).
For \texttt{GMRES<float16>} or \texttt{GMRES<int16>}, the speedup values are larger, though below the $4\times$ theoretical bound.
Again, the scaling process and memory overhead make the \texttt{GMRES<int16>} speedups inferior to the \texttt{GMRES<float16>} speedups.

\begin{figure}
  \centering
  \includegraphics[width=\columnwidth]{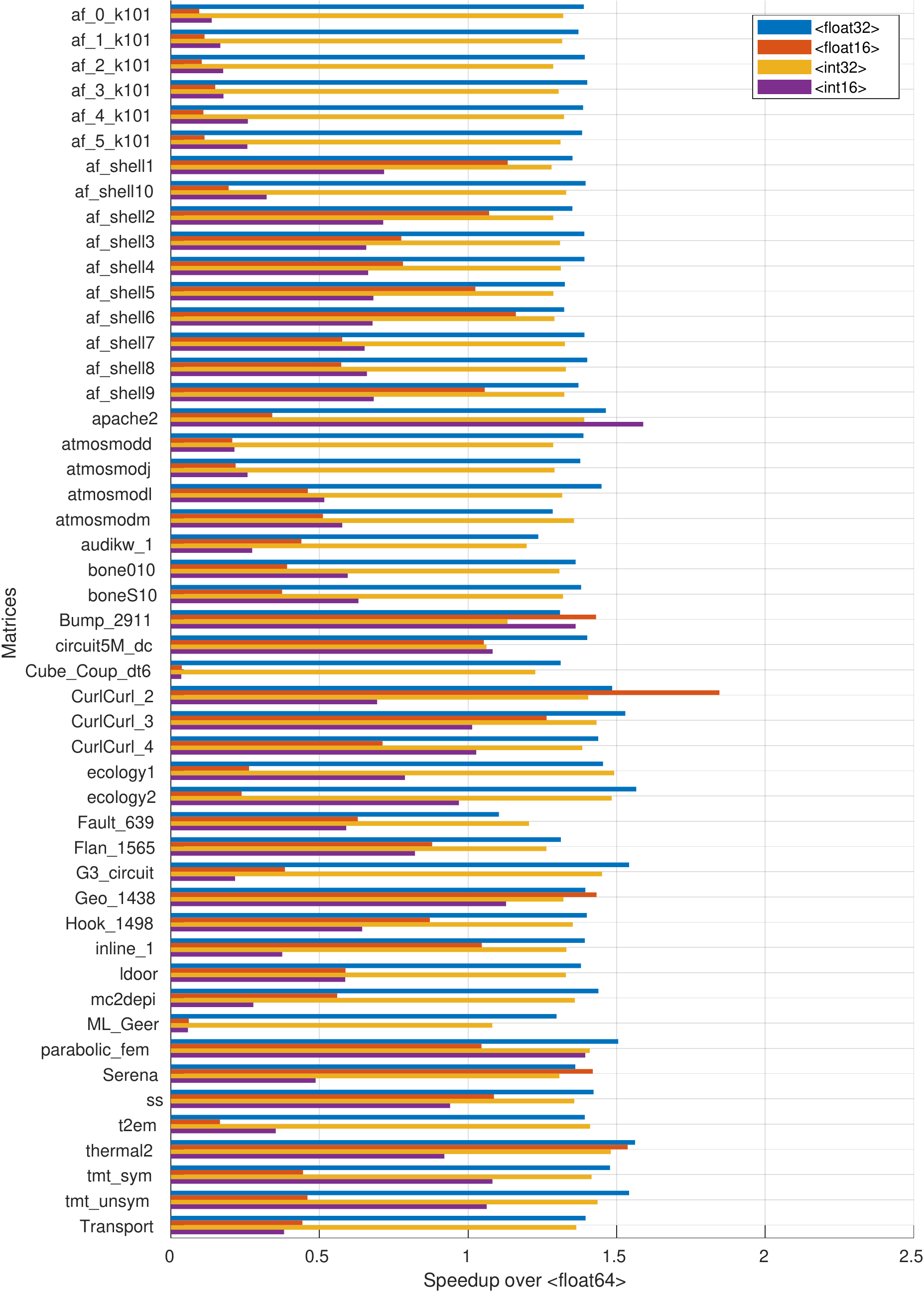}
  \caption{speedup for different CB-GMRES variants (\texttt{GMRES<precision\_format>}) over DP GMRES (\texttt{GMRES<float64>}) for increasing restart values.}
  \label{fig:performance_b4}
\end{figure}

\begin{figure}
  \centering
  \includegraphics[width=\columnwidth]{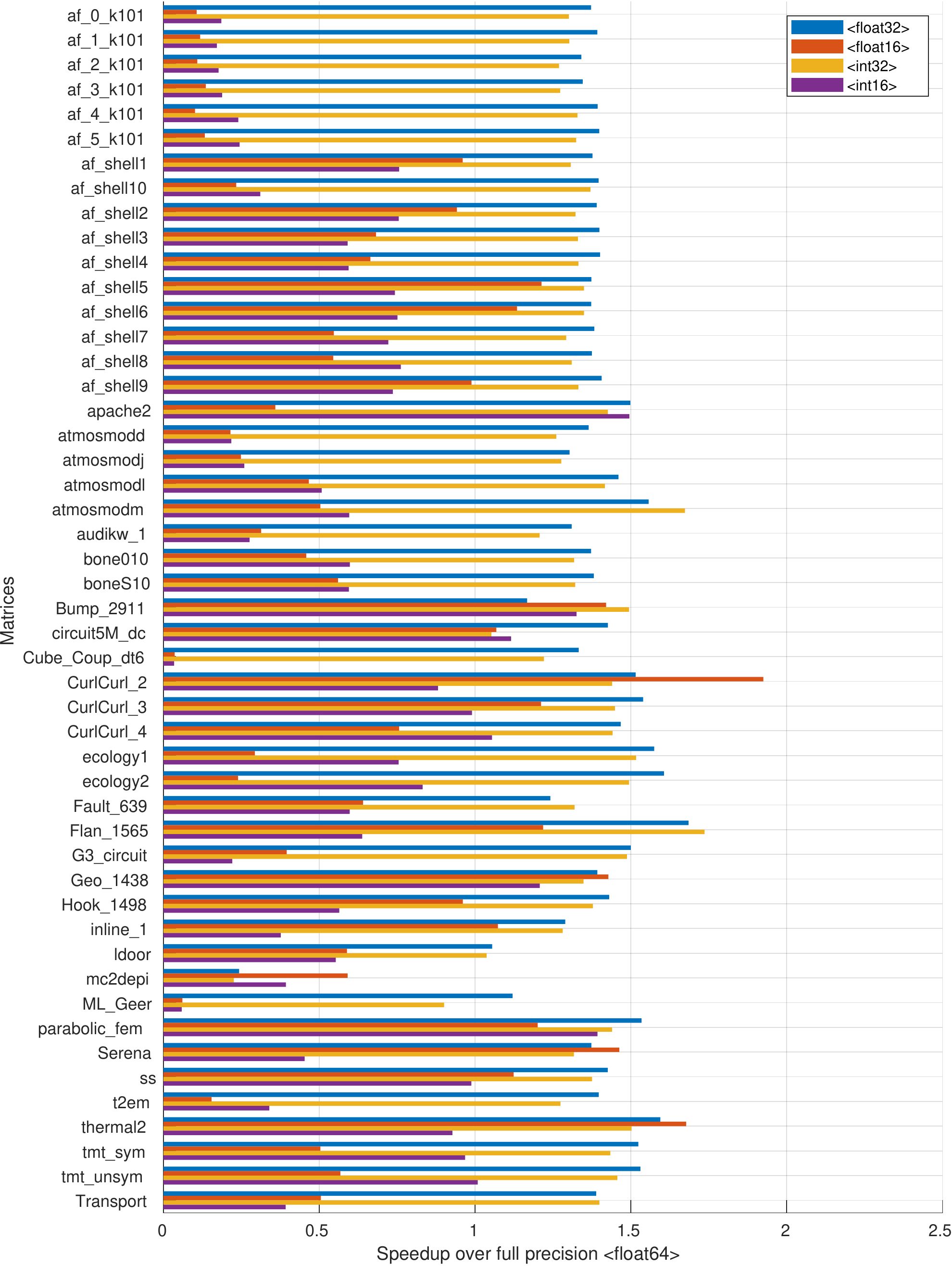}
  \caption{speedup for different CB-GMRES variants (\texttt{GMRES<precision\_format>}) over DP GMRES (\texttt{GMRES<float64>}) for increasing restart values.}
  \label{fig:bs4-speedup-adaptive}
\end{figure}

\subsection{Combining GMRES with a adaptive precision block-Jacobi preconditioner}
Finally, we investigate how the CB-GMRES interacts with a more sophisticated preconditioner and with other mixed precision techniques. For this, we switch from a scalar Jacobi preconditioner to a block-Jacobi preconditioning scheme with block-size 4, and report the performance advantages in Figure~\ref{fig:performance_b4}. As in the previous experiments, we fix the restart parameter to $m=$ 100 and run the experiments with a right-hand side vector defined by $b_i=sin(i)$, an starting guess $x_0=0$, and the residual stopping criterion set to $\|Ax^* -b\|_2\leq 10^{-9} \|b\|_2$. Compared with the results in Figure~\ref{fig:speedup}, we note a slight decrease in the speedups, which is expected as the addition of a more expensive preconditioner diminishes the performance benefits attained from storing the Krylov basis in a compressed format.

We next move from a standard block-Jacobi preconditioner to an adaptive precision block-Jacobi variant that stores the inverted diagonal blocks of the preconditioner in reduced precision if the numerical properties allow for it~\cite{anzt2019adaptive,acmtomsadaptivejacobi}. We thus combine a CB-GMRES algorithm with a multi-precision preconditioner. In Figure~\ref{fig:bs4-speedup-adaptive}, we report the speedups of CB-GMRES + adaptive precision block-Jacobi over DP GMRES + double precision block-Jacobi.
The results in that figure provide experimental evidence that the performance advantages are accumulative, and the new CB-GMRES can be efficiently combined with an independent optimization strategy targeting the preconditioner. We can naturally expect similar behaviors when combining CB-GMRES with other communication reduction techniques which target the \spmv kernel, 
or from the integration into a mixed precision iterative refinement framework.

%% file: tables/nnz.tex
af\_0\_k101 & 503,625 & 17,550,675 & 34.8 \\
af\_1\_k101 & 503,625 & 17,550,675 & 34.8 \\
af\_2\_k101 & 503,625 & 17,550,675 & 34.8 \\
af\_3\_k101 & 503,625 & 17,550,675 & 34.8 \\
af\_4\_k101 & 503,625 & 17,550,675 & 34.8 \\
af\_5\_k101 & 503,625 & 17,550,675 & 34.8 \\
af\_shell1 & 504,855 & 17,562,051 & 34.8 \\
af\_shell10 & 1,508,065 & 52,259,885 & 34.7 \\
af\_shell2 & 504,855 & 17,562,051 & 34.8 \\
af\_shell3 & 504,855 & 17,562,051 & 34.8 \\
af\_shell4 & 504,855 & 17,562,051 & 34.8 \\
af\_shell5 & 504,855 & 17,579,155 & 34.8 \\
af\_shell6 & 504,855 & 17,579,155 & 34.8 \\
af\_shell7 & 504,855 & 17,579,155 & 34.8 \\
af\_shell8 & 504,855 & 17,579,155 & 34.8 \\
af\_shell9 & 504,855 & 17,588,845 & 34.8 \\
apache2 & 715,176 & 4,817,870 & 6.7 \\
atmosmodd & 1,270,432 & 8,814,880 & 6.9 \\
atmosmodj & 1,270,432 & 8,814,880 & 6.9 \\
atmosmodl & 1,489,752 & 10,319,760 & 6.9 \\
atmosmodm & 1,489,752 & 10,319,760 & 6.9 \\
audikw\_1 & 943,695 & 77,651,847 & 82.3 \\
bone010 & 986,703 & 47,851,783 & 48.5 \\
boneS10 & 914,898 & 40,878,708 & 44.7 \\
Bump\_2911 & 2,911,419 & 127,729,899 & 43.9 \\
circuit5M\_dc & 3,523,317 & 14,865,409 & 4.2 \\
Cube\_Coup\_dt6 & 2,164,760 & 124,406,070 & 57.5 \\
CurlCurl\_2 & 806,529 & 8,921,789 & 11.1 \\
CurlCurl\_3 & 1,219,574 & 13,544,618 & 11.1 \\
CurlCurl\_4 & 2,380,515 & 26,515,867 & 11.1 \\
ecology1 & 1,000,000 & 4,996,000 & 5.0 \\
ecology2 & 999,999 & 4,995,991 & 5.0 \\
Fault\_639 & 638,802 & 27,245,944 & 42.7 \\
Flan\_1565 & 1,564,794 & 114,165,372 & 73.0 \\
G3\_circuit & 1,585,478 & 7,660,826 & 4.8 \\
Geo\_1438 & 1,437,960 & 60,236,322 & 41.9 \\
Hook\_1498 & 1,498,023 & 59,374,451 & 39.6 \\
inline\_1 & 503,712 & 36,816,170 & 73.1 \\
ldoor & 952,203 & 42,493,817 & 44.6 \\
mc2depi & 525,825 & 2,100,225 & 4.0 \\
ML\_Geer & 1,504,002 & 110,686,677 & 73.6 \\
parabolic\_fem & 525,825 & 3,674,625 & 7.0 \\
Serena & 1,391,349 & 64,131,971 & 46.1 \\
ss & 1,652,680 & 34,753,577 & 21.0 \\
t2em & 921,632 & 4,590,832 & 5.0 \\
thermal2 & 1,228,045 & 8,580,313 & 7.0 \\
tmt\_sym & 726,713 & 5,080,961 & 7.0 \\
tmt\_unsym & 917,825 & 4,584,801 & 5.0 \\
Transport & 1,602,111 & 23,487,281 & 14.7 \\

%% file: s6-conclusions.tex
\section{Summary and Outlook}
\label{sec:conclusion}

We have introduced and evaluated a communi\-cation-reduction version of GMRES that maintains the orthogonal basis in a 
compressed (compact) form  while performing all arithmetic in double precision.
The combination of these two factors aims to reduce
the traffic between memory and the processor arithmetic units while maintaining the accuracy of the search
directions generated during the optimization process and extracting the performance from hardware-supported arithmetic.
In contrast, the memory storage provides (to a certain extent) enough flexibility to evaluate distinct 16-bit and 32-bit formats, including floating point and fixed point variants.

We have integrated a high-performance realization of the GMRES with compressed orthogonal basis into the Ginkgo
framework for sparse linear systems. The performance evaluation of this solver on a recent NVIDIA V100 GPU demonstrates the practical advantages of the communication-reduction technique, which is aligned with the acceleration
that could be expected from Amdahl's law. 
On the one hand, the speedups are more notable for the 32-bit floating point 
format, followed closely by its 32-bit fixed point counterpart. On the other hand, the 16-bit formats further reduce the 
communication volume, but they regularly fail to preserve the convergence characteristics of the GMRES solver.
Overall, we believe that the proposed technique is useful as it tackles the memory wall problem that is present in
current processors. Furthermore, its benefits are orthogonal and, therefore accumulative, 
to those that can be attained with other communication-reduction
techniques applied, for example, to the preconditioner, the realization of \spmv, or the GMRES algorithm itself.

In future work we will investigate whether compression techniques that are traditionally used to compress large datasets can pose an alternative to the use of of 32-bit and 16-bit fixed and floating point formats to store the compressed basis vectors.